# Exuberant innovation:
# The Human Genome Project

Monika Gisler[1], Didier Sornette[2,3] and Ryan Woodard[2]

[1]ETH Zurich, D-ERDW, NO, CH-8092 Zürich, +41 78 919 5058
monikabgisler@gmail.com (corresponding author)
[2]ETH Zurich, D-MTEC, Chair of Entrepreneurial Risks, Kreuzplatz 5, CH-8032 Zurich
dsornette@ethz.ch; rwoodard@ethz.ch
[3]Swiss Finance Institute, University of Geneva, 40 blvd. du Pont d'Arve, CH-1211 Geneva 4

"And all this back and forthing over who did what and what strategy was used and which money was public and which was private is probably going to sink below the radar screen." (Francis Collins)[1]
"The prevailing view is that the genome is going to revolutionize biology, but in some way, it's overhyped. In the end, the real insights are coming from individuals studying one gene at a time in real depth." (Gerald Rubin)[2]

**Abstract**

We present a detailed synthesis of the development of the Human Genome Project (HGP) from 1986 to 2003 in order to test the "social bubble" hypothesis that strong social interactions between enthusiastic supporters of the HGP weaved a network of reinforcing feedbacks that led to a widespread endorsement and extraordinary commitment by those involved in the project, beyond what would be rationalized by a standard cost-benefit analysis in the presence of extraordinary uncertainties and risks. The vigorous competition and race between the initially public project and several private initiatives is argued to support the social bubble hypothesis. We also present quantitative analyses of the concomitant financial bubble concentrated on the biotech sector. Confirmation of this hypothesis is offered by the present consensus that it will take decades to exploit the fruits of the HGP, via a slow and arduous process aiming at disentangling the extraordinary complexity of the human complex body. The HGP has ushered other initiatives, based on the recognition that there is much that genomics cannot do, and that "the future belongs to proteomics". We present evidence that the competition between the public and private sector actually played in favor of the former, since its financial burden as well as its horizon was significantly reduced (for a long time against its will) by the active role of the later. This suggests that governments can take advantage of the social bubble mechanism to catalyze long-term investments by the private sector, which would not otherwise be supported.

O43 - Institutions and Growth
G12 - Asset Pricing; Trading volume; Bond Interest Rates

---

[1] Francis Collins, interview with Leslie Roberts, 19 August 1999; Roberts et al. 2001.
[2] Gerald Rubins, interview with Elizabeth Pennisi, Februrary 2000; Roberts et al. 2001.

# 1–Introduction

The Human Genome Project (HGP), a genuine innovation in the molecular biology sector, begun formally in 1990. It was coordinated by the U.S. Department of Energy and the National Institutes of Health, and was completed in 2003. It was one of the largest international scientific research projects, with the primary goal of determining the sequence of chemical base pairs which make up DNA, and to identify and map the approximately 20,000–25,000 genes of the human genome from both a physical and functional standpoint (Watson and Cook-Deegan, 1991; Cook Deegan, 1991; 1994; Gilbert, 1992; Hilgartner, 1994; 1997; 1998; 2004; Koonin, 1998; Jordan and Lynch, 1998; Roberts, et al., 2001; Kieff, 2003). It was launched on the rational that, with all the genes identified and available in computerized data banks, genetic mapping[3] and sequencing data would utterly transform biology, biotechnology, and medicine in the next century.

This large-scale project provides an excellent example to study the patterns of an innovation at large and the social bubble hypothesis that we have formulated elsewhere (Gisler and Sornette, 2009; Sornette and Gisler, forthcoming). We hypothesized that, when new technology or scientific options open up, and individual or groups believe to be ready for it, then they dive into these new opportunities, often without apparent concern for the risks and possible adverse consequences. According to the social bubble hypothesis, the social interactions between enthusiastic supporters weave a network of reinforcing feedbacks that lead to widespread endorsement and extraordinary commitment by those involved in the project. The term "bubble" is borrowed from the financial economic literature, in which a bubble is defined as a transient appreciation of prices above fundamental value, resulting from excessive expectations of future capital gain.

Sornette (2008) has suggested that the following major inventions could be instances of such social bubbles: the great boom of railway in Britain in the 1840s, the Human Genome project, the cloning of mammals (Dolly, the sheep), and the ICT (Internet-Communication-Technology) bubble culminating in 2000. The adventure of nanotechnology or the craze over Haute Couture ('the democratization of fashion design') could be added to the list. A property shared by these cases is that they were all characterized by extremely high expectations concerning the outcome of the proposed research and/or innovation project. Therefore, enthusiasm was high at the start of the project, leading to the readiness to take large risks, which may have been at the origin of innovations that can turn out to be tremendously valuable on the long term. Some of these innovations led to fast societal progress and structural changes, others captured the imagination of large groups and proceeded along a roller-coaster of rising expectations, steep growth and spectacular downturns, with potential future benefits still uncertain. They all constitute an essential element in the dynamics of important inventions, and are thus crucial for society.

Our working hypothesis is that the nucleation and growth of bubbles play a key role in reducing collective risk aversion that normally restrict innovation and discovery processes. Innovation is understood as a social process that brings together various actors of different backgrounds and interests. During its development, the process is framed by interactions and novel relationships among science, business/industry, economy, and politics. Because technological change is so vital for long-run economic growth, it is of fundamental importance to understand how individuals, firms, and the public (via the government) obtain the resources needed to undertake their investments in innovation and invention. It is also important to understand how the availability of such resources, including the manner in which they are accessed as well as the amounts that can be raised, influences the rate, direction, and organization of technological development. The present paper intends to cast light on these issues by a detailed analysis of the development of the Human Genome Project (HGP), which constitutes a paradigm of a crucial technology jump via a large-scale research project. In particular, we focus on the question of how the HGP was funded during its lifetime, as this provides an objective and precise metric reflecting the choices of and conflicts between the different involved parties.

---

[3] Genome mapping is the creation of a genetic map assigning DNA fragments to chromosomes.

To put things into perspective, recall that the costs of the HGP, estimated early on at about $3 billion, has engendered great concern, raising fears about "big science" and the effect that a project of this magnitude might have on other areas of biological research (DeLisi, 1988; Roberts, 1990/248). This figure of $3 billion has come up in an early discussion of whether or not to sequence the Human Genome; it refers to the numbers calculated by Walter Gilbert (*1932), an early defender of the Human Genome Project. At a meeting at Cold Spring Harbor Laboratory in June 1986, several of the most famous scientists in molecular biology discussed the cost of a potential sequencing process of the human genome (Lewin, 1986b/233). Gilbert counted that at $1 per base pair, a reference sequence of the human genome could be obtained for about $3 billion. This cost projection provoked uproar, so that in response Gilbert suggested to first concentrate on the 1 percent of the genome containing biologically known function, then to do the next 10 percent, and only afterwards finish the job, devoting equal resources to each phase (Cook-Deegan, 1994). Another attempt to estimate the cost of the whole HGP was prepared by the NRC committee in 1988, which projected the need for $200 million per year over 15 years, to support research centers, grants and technology development, and administration (Cook-Deegan, 1991). While these two estimations end up given the same total amount, they refer to very different itemizations of the costs of the HGP, which illustrate the difficulties in estimating its global costs, especially at its inception (Roberts, 1987b/237).

However, it was argued in many places that the costs should not be considered so extravagant because, in addition to the sequencing of the human genome, they would also cover a wide range of other scientific activities extending over a 14-year period (1990–2003) including studies of human diseases, of experimental organisms (such as bacteria, yeast, worms, flies, and mice), the development of new technologies for biological and medical research, computational methods to analyze genomes, and investigations on ethical, legal, and social issues related to genetics. Human genome sequencing was argued to represent just a fraction of the overall budget.

The issue of the funding of the HGP is made more intricate and interesting by the fact that, in addition to the major contributions of the U.S. government (see table 1) and of other public institutions (the UK Wellcome Trust, and other countries such as France, Germany, Japan, and so on), Celera Genomics and other firms were pursuing separately the route of private venture capital. Yet, the issue of where the money came from, how it was raised, what arguments were brought forth in favor or against the HGP, has hardly been scrutinized in the literature. In our study, we will rely in particular on the data collection performed by a group organized by Robert Cook-Deegan on the money spent for biotechnology endeavors (Stanford-in-Washington. World Survey of Genomics Research www.stanford.edu/class/siw198q/websites/genomics/entry.htm; retrieved June 1, 2009; Cook-Deegan et al., 2000; Reineke and Cook-Deegan, 2008; Chandrasekharan et al., 2009).

In this paper, we focus on the question of investment associated with the HGP. We try to understand the influences that motivated to invest into the project, and how its directions were channeled. This allows us to assess and quantify the importance of government spending and funding (and its legitimization), based on the premise by Nelson (1959) that public subsidy of science is legitimized by the recognition of inefficiencies in the market for scientific knowledge. Nelson contents that the uncertain nature of the output of basic research means that private investors cannot be sure they will benefit from their investment and, as a consequence, a purely market-based system would tend to invest at lower than the economically and socially desirable levels. Because of the uncertainty about the direction of any future development of basic research, private companies might be drawn to withdraw from any kind of funding. However, in the presence of public investment on basic research, private investors could come in later, at a lower risk level (Fabrizio and Mowery, 2007). This raises the interesting question of what percentage of the GDP of a country should be allocated to science by governments. Determining the right level is very difficult given the large uncertainties (Sornette and Zajdenweber, 1999). Too little would thwart innovation, productivity and economic growth. Too much could be a waste (Gersbach et al., 2008).

This discussion can be understood within the context of the Government-Industry-University relationship, recently labeled as the "Triple Helix": the increased importance of knowledge gives universities a central position in the transfer of academic knowledge to foster industrial innovations according to a back-and-forth interactive and iterative process, rather than via the obsolete linear model of innovation (Leydesdorff and Etzkowitz, 1998; Etzkowitz 2002; 2003; see also Stokes, 1997). As firms raise their technological level, they move closer to an academic model, engaging in higher levels of training and in sharing of knowledge. Governments act as a public entrepreneur and venture capitalist in addition to their traditional regulatory role in setting the rules (Mansfield, 1995). Moving beyond product development, innovation becomes an endogenous process, encouraging hybridization among the institutional spheres.

We thus ask the question whether or not this "Triple Helix" mechanism was at play for the HGP. Implicit in this question is the assumption that, generally speaking, governments fund science, whereas technology is mainly privately funded. Given that governments support by various means the generation of inventions with the goal of increasing productivity (Orsenigo, 1993), a very important question in this respect is: who was first? Was it the private sector (drug companies and/or venture capitalists) that convinced the government to step in and provide coordination help and funding? Or was it a set of universities? Or was it the government which more or less on its own appropriated a fancy project to make "a big show" – comparable to the Apollo Program (Gisler and Sornette, 2009)? For the sake of conciseness, we will focus on the U.S. only, even though the UK, namely the Sanger Institute, funded by the Wellcome Trust, was apparently an equally important partner of the HGP (Balmer, 1996; Sulston and Ferry, 2002). In this respect, we have analyzed the several progress reports of the different agencies involved. Furthermore we have scrutinized dozens of scientific journals as well as monographs based on interviews with the protagonists. We will investigate whether standard cost-benefit and portfolio analysis can explain the HGP. This will allow us to test our hypothesis that, with over-optimistic expectations, people focus almost solely on the expected returns of an invention and tend to forget its risks. There are risks whose magnitude are so big that they cannot be funded by private investors, thus only governments can take the 'systemic' large risks on their shoulders, by using the largest reservoir of funds provided by the pool of taxpayers.

We structure the paper as follows. Section 2 describes the context and process of the nucleation of the HGP. Sections 3–7 document and compare the public component of the HGP to the private initiatives, their interplay, rivalry and entanglement, which were defining sociological characteristics of its development. Section 3 describes an early attempt by Walter Gilbert to take the genome project private. Section 4 presents the public approach to the HGP. Section 5 discusses Craig Venter's Celera Genomics approach to the human project and how it forced the publicly funded project to evolve. Section 6 describes the development and progressive transformation of the public effort. Section 7 shows how the rivalry between the public and private projects led to the transformation of the former to endorse more risky approaches with stronger links with industry. This section also describes the explosion of investments in new biotech firms by venture capitalists and by Wall Street investors in the second half of the 1990s. Section 8 presents quantitative evidence for the existence of a financial biotech bubble, paralleling the development of the HGP, which culminated in March 2000, a few month before the official announcement of the completion of the HGP on June 26, 2000. We take special care in trying to separate the pure biotech component of the bubble from the overall ICT bubble that also crashed in March 2000. Section 9 describes the completion of the HGP. Section 10 concludes that the public-private competition documented here provides support to the social bubble view of the nucleation, development and completion of the HGP. We also briefly comment on how much of the expectations justifying the HGP have been reassessed ex-post, further supporting the existence of exuberant over-optimism characteristic of a bubble Zeitgeist.

**2–Emergence**

The Human Genome Project (HGP) is an assemblage rather than a single entity, which emerged from decades of research on genetics (Cantor, 1990; Hedgecoe and Martin, 2008; Rheinberger, 2008). The scientific foundation for a human genome initiative existed at the U.S. national laboratories before the establishment of the first genome project. Besides expertise in a number of areas critical to genomic research, the laboratories had a long history of conducting large multidisciplinary projects. The HGP's actual start is difficult to define. In general, a few important workshops in the nascent field of genome analysis in the 1980s are seen as the beginning of the initiative. Particularly, an Alta summit in 1984 (Cook-Deegan, 1989) and a Santa Cruz meeting in 1985 (Sinsheimer, 1989) are seen as the launch of the Human Genome initiative. At the latter, the attending group of high-ranking scientists from the U.S. and the UK decided that it made sense to develop systematically a genetic linkage map, i.e. a physical map of ordered clones. Once a gene has been isolated, the next step is to sequence it, that is, to determine its internal structure. The sequencing efforts, the panel agreed, should first focus on automation and development of faster and cheaper techniques (Cook Deegan, 1994).

Besides the point that the best possible investment a nation can make for its future was, next to education, science, the main argument at that time was that investing into genome research was investing into research on the genes believed to be involved in the diseases and on their potential cure. In fact, research on genes was fueled by the aspiration to track down diseases, assumed to be inherited via genes (Cook-Deegan, 1994; DeLisi, 1988). For example, when the Cystic Fibrosis gene was found in 1989, researchers were certain that therapy was around the corner. The same is true for the many cancer types, or hereditary diseases. Knowledge of the genome and availability of probes for any gene was seen as crucial for the progress on diagnosis and therapeutics. In a seminal paper, e.g., Renato Dulbecco urged for including the study of the cellular genome in order to progress on cancer research (Dulbecco, 1986). Moreover, the very positive reception of the government and the public to the HGP was most likely due to the alleged priority set forth to detect disease genes. When the project started, though, other topics were given priority over finding genes that might hold diseases. And two decades later, one has to state that there is still much to do, and one is in fact tempted to declare that "The disease has contributed much more to science [i.e. to the support for the Human Genome Project] than science has contributed to the disease." (Jack Riordan, cited by Pearson, 2009/460: 165).

To DeLisi, then head of the Office of Health and Environmental Research at the Department of Energy (DOE), the genome project was a logical outgrowth of DOE's mandate to study the effects of radiation on human health. At his urging, Los Alamos National Laboratory hosted a workshop in Santa Fe, New Mexico, in March 1986, the first workshop under the auspices of DOE (DeLisi, 1988; 2008). The idea laid out at this Santa Fe workshop quickly gained momentum, dominating discussion at a meeting a few months later at the Cold Spring Harbor Laboratory in New York. By then, biologists were beginning to think the project just might be doable (DeLisi, 1988; Cook-Deegan, 1994). Mapping the human genome seemed not to be too far, estimates varied between two and five years. The symposium marked a transition from emphasizing the sequencing of the human genome to a broader plan for genetic linkage mapping, physical mapping, and the study of nonhuman organisms. It is worthwhile to mention that, according to Cook-Deegan (1991), the public remained largely ignorant of the project even after it had been under way for a couple of years.

DeLisi eventually gained support for the project, first from his superiors at DOE and then from Congress, starting a small Human Genome Initiative within DOE in 1986 (Cantor, 1990). In April 1987, the initiative was endorsed by a report from the Department's Health and Environmental Research Advisory Committee (HERAC) (Subcommittee on Human Genome of the Health and Environmental Research Advisory Committee for the U.S. Department of Energy Office of Energy Research Office of Health and Environmental Research, 1987). The HERAC report urged DOE and the nation to commit to a large, long-term, multidisciplinary, technological undertaking to order and sequence the human genome (Palca, 1987; Barnhart, 1989). The physical map of the human genome – the report presumed – could be done by DOE as well as universities and industry.

Involvement in this initiative was seen as a consequence of DOE's demonstrated expertise in handling projects of this size and scope. Subsequent reports from the National Academy of Sciences and the Congressional Office of Technology Assessment (OTA) supported the HERAC report by endorsing a major national effort at a sustained level of $200 million annually (U.S. Congress, Office of Technology Assessment, 1988). The initiative was seen as having substantive long-term impacts on basic science and on biotechnology and pharmaceutical industries, as well as on the practice of medicine. The long-range goal of this dedicated research was to develop and provide the broad array of resources and technologies that would allow the complete characterization of the human genome at the molecular level. The OTA report differs from others on the topic in that it explored implications of

the project that had been neglected thus far. Especially, it pointed out that those physical maps, genetic linkage maps, clone repositories and genetic databases will all become available as a result of the project, and will have immediate utility for the biological sciences. This report also raised some difficult issues relating to patents, copyrights and technology transfer that will arise as private companies and foreign governments join federally-supported research laboratories in working on the project.

Today, DOE is seen as the first federal agency to have announced and funded a genome program (see www.ornl.gov/sci/techresources/Human_Genome/project/whydoe.shtml; retrieved March 3, 2010). The fact, however, that DOE was lobbying for the project only heightened some biologists' unease, who put emphasis on the peer-review system. Little enthusiasm came from younger scientists, who feared that a mega billion-dollar project would divert money away from single investigator-initiated research grants and slow down the pace at which high-quality biological and medical research was carried out in the U.S. (Lewin, 1986c/233; 1986d/233; Watson, 1990). The fear furthermore was not so much about big science, but alleged bad science (DeLisi, 1988; Roberts, 2001/291). In fact, the Human Genome Project was never considered as passing for big science (in reports such as the OTA report e.g.), and it was never perceived as big science by Science and Technology Studies scholars, even though it was at the center of biomedical research after 1987 and retained this status for several years. At the same time, the project was never really put into question. Fears were more concerned with its size, and whether or not it would drain money from other biotechnology projects (Roberts, 1990/248). Back in 1986, it is noteworthy that the interest in sequencing the entire human genome was sometimes decreasing with more enthusiasm was placed on mapping (Lewin, 1986d/233).

Political posturing continued until 1988, when a National Research Council (NRC) committee gave the project its official seal of approval (Committee on Mapping and Sequencing the Human Genome, National Research Council, 1988). It was urged that federal funding should rise quickly to $200 million a year, with the project planned to be completed in approximately 15 years (Watson, 1990; Roberts, 1988a/239). At the same time, an ad hoc advisory committee on complex genomes within NIH followed Wyngaarden's proposal to establish an Office of Human Genome Research to be headed by a new associate. In late 1989, the Human Genome Project began to consolidate. In October 1989, under James Watson, the Office of Human Genome Research became the National Center for Human Genome Research (NCHGR) (Roberts, 1988b/241; Roberts, 1989b/245). Watson declared the official start of the genome project as October 1990, corresponding to the beginning of fiscal year 1991. If there was initially uncertainty over how the NIH and DOE program would be coordinated (Watson, 1990), with this move, NIH was firmly established as the lead agency. The project was urged to start by constructing maps of the human chromosomes. Full-scale sequencing would be postponed until new technologies made it faster and cheaper (Cook-Deegan, 1994; Roberts, 2001/291). Altogether, it had taken five years for the genome project to be translated from an idea into the beginnings of an international scientific project (Watson and Cook-Deegan, 1991).

### 3–Private I: Walter Gilbert goes private

Nobel laureate Walter Gilbert, a molecular biologist who worked with James Watson in the early 1960s at Harvard, became impatient with the cautious approach to sequencing. Perhaps useful to understand his state of mind, he is known to have compared the research of the human genome to the search of the *Holy Grail*. Arguing that the technology was already good enough to sequence the human genome, he decided to take the genome project private. In 1986, he left the National Research Council, announcing the launch of a genome company, called *Genome Corporation* (Palca, 1987; Roberts, 1987c/237; 2001/291; Cook-Deegan, 1994). His intention was to "create a catalog of all human genes which would be made available to everyone for a price." (Walter Gilbert, cited in Roberts, 1987c/237: 358). He expected that customers would include the academic research community as well as the pharmaceutical industry. By this statement, he provoked a major controversy. The concerns involved the possibility that exchange of data between scientists would be slowed down or barred entirely, and furthermore that access to some data would be locked out. The expectation of having to compete with corporate scientists left many academics uneasy (Roberts, 1987c/237; Sulston and Ferry, 2002). Robert Cook-Deegan, then with the Office of Technology Assessment (OTA), expressed loudly what many may have been thinking: "If a company behaves in what scientists believe is a socially responsible manner, they can't make a profit." (Robert Cook-Deegan, cited in Roberts 1987c/237).

According to an interview of Gilbert with Cook-Deegan, Gilbert's idea for Genome Corporation was to construct a physical map, do systematic sequencing, and establish a database (Cook-Deegan, 1994, interview with Gilbert). He did not speculate publicly on how long his mapping and sequencing effort would take, but admitted that his time table was generally more aggressive than that of other people. He also reckoned the entire sequencing effort to cost far less than the DOE estimate, more like $300 million, and to be accomplished within a decade by a modestly sized private company (Palca, 1987).

His plan, which was remarkably similar to J. Craig Venter's vision half a decade later (see below), was to set up a sequencing factory to churn out the data, which he intended to copyright and sell. This included selling clones from the map, serving as a sequencing service, and charging user fees for access to the database. The market would be academic laboratories and industrial firms, such as pharmaceutical companies, that would purchase materials and services from Genome Corporation. The purpose was not so much to do things that others could not, but to do them more efficiently, so that outside laboratories could purchase services more economically than performing the services themselves. These premises fueled Gilbert's quest to find funding from venture capitalists over the course of 1987 and into 1988. In January 1987, he was approached by a foundation in order to help create such an institute. The idea died after the foundation funded a study to assess the genome project at the National Research Council of the National Academy of Sciences. Moreover, by late 1987, Wall Street's enthusiasm for biotechnology had turned into skepticism, and the stock market crash in October made capitalizing Genome Corporation impossible. The highly publicized efforts to start a genome project by the federal government made prospective investors distrustful of competing with the public domain. Genome Corporation could succeed only if Gilbert stayed so far ahead of academic competition that others would come to him for services, rather than waiting for the information and materials to be made freely available. With the failure of the efforts to raise sufficient funds, Gilbert's venture died, and with it – at least for some time – the feud between public and private teams (Roberts, 1987b/237; 2001/291).

**4–Public I**

At the beginning of 1987, when Gilbert formulated his plans for Genome Corporation, there was no center to support efforts in genome mapping and sequencing. Two federal agencies emerged eventually that competed for the leadership of the genome project (Roberts, 1987a/237; 1988a/239). In 1988, the National Institutes of Health (NIH) and the Department of Energy (DOE) eventually signed a Memorandum of Understanding to facilitate cooperation and coordination of genome research and development and to establish a joint advisory committee to coordinate these activities. The memorandum also established an interagency working group in which staff members of NIH and DOE met regularly to discuss research of mutual interests, as well as agency priorities.

In April 1990, NIH and DOE published a five-year plan, whose goals included the completion of a genetic map, a physical map, and the sequence of model organisms by 2005 (U.S. Department of Health and Human Services and U.S. Department of Energy, 1990). In October 1990, the start of the project was officially announced. And by the end of the year, both the Department of Energy and the National Institutes of Health had genome programs with budgets totaling almost $84 million, and similar dedicated genome programs were launched in the United Kingdom, Italy, the Soviet Union, Japan, France, and the European Communities (Watson, 1990; Watson and Cook-Deegan, 1991; Cook-Deegan, 1994; for France see esp. Kaufmann, 2004).

Notwithstanding a neoliberal orientation of the policies of the 1980s, the U.S. governmental expenditures for research did not decrease; on the contrary, they increased annually (Barben, 2007). A body which tremendously gained from this process was the National Institutes of Health (NIH), an institution responsible for biomedical and health related research, and a part of the U.S. Department of Health and Human Services.

As the genome project gained congressional funding and scientific respectability, NIH seized control from DOE. NIH director James Wyngaarden announced that they would create a special office for genome research. The project was initially headed by James D. Watson (*1928), the American molecular biologist, best known for being one of the discoverers of the structure of DNA. With this initiative, NIH was firmly established as the lead agency (Cook-Degan, 1994; Watson, 1990; Roberts, 1988b/241). It has remained so, even as the project gathered international collaborators and Britain's Wellcome Trust took on a prominent role in 1992. Watson proved a shrewd strategist: Knowing that Congress did not have the patience to wait 15 years for results, he relentlessly pushed forward the first stage of the project and its most tangible goal, the build-up of maps of human chromosomes (Roberts, 2001/291). Even though disease genes captured the public imagination and kept the dollars flowing, it was Watson's (and others') vision that the project would begin with genetic and physical mapping and gradually develop technology to sequence the whole genome, in order to "to find out what being human is." (Roberts, 1989a/243: 167). He predicted that a detailed genetic map of all the human chromosomes would be finished within five years.

The issue of gene patenting led to a change of leadership. Quarrelling over patenting was largely triggered – even though not new – by J. Craig Venter, then NIH biologist, who in July 1991 announced that NIH was filing patent applications on thousands of partial genes. Even though the series of questions Venter opened up could be considered *a priori* as legitimate, the issue of patenting as a turning point in the commercialization of molecular biology caused controversy (Smith Hughes, 2001; Sulston and Ferry, 2002; Shreeve, 2004). Patenting driven by profit motives was deeply repugnant to Watson. He felt strongly that the sequence data flowing from the HGP should remain within the public domain, freely available to all. Meeting opposition on his view, he stepped down from his position as director of the NIH-sponsored project in 1992 (Roberts, 1992/256). He was replaced by Francis Collins in April 1993 (Roberts, 1993/262). In 1997, the name of the Center changed to *National Human Genome Research Institute* (NHGRI) (Cook-Deegan, 1989; Barnhart, 1989).

**5–Private II: Venter-ing**

Craig Venter was not only the initiator of the discussion on patenting genes, he was also pivotal when it came to commercialize genome research. A scientist at the NIH during the early 1990's, running a large sequencing lab at the National Institute for Neurological Disorders and Stroke, he felt in 1991 that private companies could sequence genomes faster than publicly funded laboratories (Shreeve, 2004). Venter boasted that a newly developed approach could do the sequencing better, and for a fraction of the costs the official Human Genome Project was budgeting. Venter claimed to be able to find 80% to 90% of the genes within a few years only (Adams et al., 1991; Roberts, 1991/252). This ushered the era of competition between the public and the private initiatives in terms of speed and efficiency.

Following his vision, Venter left the NIH in 1992 to set up his own biotechnology company. In complete contrast to the failure of Walter Gilbert' attempts to garner private funds in 1987, the time now seemed ripe for the development of genetic research by the private sector. Venter set up The Institute for Genomic Research (TIGR), a non-profit firm, funded by the investment company HealthCare Investment Corporation. He was being offered $70 million to try out his own gene identification strategy. In addition, TIGR was also one of the six centers receiving support from the NIH (Shreeve, 2004). The arrangement was that its sister company, Human Genome Sciences (HGS), led by William Haseltine, would commercialize the products developed by TIGR. The deal was that HGS should have exclusive access to TIGR's Expressed Sequence Tags (EST's) for a certain time

before publication.[4] Academic scientists would be able to look at the TIGR database freely after that, but the commercial company would have access rights to any further commercial developments. The company sold an exclusive license for prior access to the information to the pharmaceutical giant SmithKline Beecham for $125 million (Sulston and Ferry, 2002).

In 1995, TIGR published the first completely sequenced genome, that of the bacterium *Haemophilus influenza*. The scientists had carried it out in just a year, using a riskier technique, called *whole genome shotgun sequencing*, that NIH had insisted wouldn't work and wouldn't fund (Sulston and Ferry, 2002). Sequencers in the publicly funded project had adopted a conservative, methodical approach, starting with relatively small chunks of DNA whose positions on the chromosome were known, breaking them into pieces, then randomly selecting and sequencing those pieces and finally reassembling them (Bostanci, 2004). In contrast, Venter simply shredded the entire genome into small fragments and used a computer to reassemble the sequenced pieces by looking for overlapping ends (Roberts, 2001/291).

Among other biotech firms involved in gene sequencing, *Celera Genomics* was founded in 1998 by Venter in conjunction with the Perkin-Elmer Corporation, the manufacturer of the world's fastest automatic DNA sequencers. The company would single-handedly sequence the entire human genome in just three years, they announced, and for a mere $300 million (Marshall and Pennisi, 1998/280). Venter's goal was to privately sequence the human genome in direct competition with the public efforts supported by the NIH and DOE and by the governments of several foreign countries (Venter in Science, 1998/280: 1540–1542); see also Human Genome News 1998/9/3; 2000/11/1–2). Using 300 Perkin-Elmer automatic DNA sequencers along with one of the world's most powerful computers, Celera sequenced the genomes of several model organisms with remarkable speed. Venter called his effort "a bargain by comparison to the genome project." (Roberts, 1991/252: 1619). They continued in their efforts of sequencing the entire human genome at a cost of a few million dollars per year, instead of the hundreds of millions of the public project. Leaders of the latter began to worry: Should Congress fell for Venter's boldness, it might pull the plug off the public project. His plan would never work, they countered, and the sequence would be riddled with holes and impossible to reassemble (Roberts, 2001/291).

In a crucial test of the shotgun strategy, Celera first tackled the 180-megabase genome of the fruit fly *Drosophila melanogaster* (Butler, 1999/401; Pennisi, 2000/287). Venter teamed up with a publicly funded team headed by Gerald Rubin of UC Berkeley, and by September 1999 announced to have carried it out (Shreeve, 2004). Although this did not mean that they had fully finished or even assembled the 180 megabase sequence, they had run enough samples through the machines to cover the whole genome. According to Venter, this addressed the criticism raised by the public genome project and proved that the shotgun methods could work on a big, complex genome.

Venter was thus in the position to threaten the fragile alliance among the publicly funded sequencing labs. The contest was punctuated by dueling press releases (Sulston and Ferry, 2002). First Venter announced in October 1999 that his crew had sequenced one billion bases of the human genome, a feat rejected by the HGP, which noted that Celera hadn't released the data for other researchers to check. Then NIH jumped into the game, announcing in November that it had completed 1 billion bases. Venter countered in January 2000 that his crew had compiled DNA sequence covering 90% of the human genome; the public consortium asserted in March that it had completed two billion bases, and so on. Issues of data access heated up too, with the public consortium denouncing Venter for his plan to release his data on the Celera Web site rather than in GenBank, the public database.[5] The feud became increasingly ugly, with each side disparaging the other's work and credibility in the press (Sulston and Ferry, 2002).

---

[4] An Expressed Sequence Tag (EST) is a tiny portion of an entire gene that can be used to help identify unknown genes and to map their positions within a genome. ESTs provide researchers with a quick and inexpensive route for discovering new genes, for obtaining data on gene expression and regulation, and for constructing genome maps; (http://www.ncbi.nlm.nih.gov/), retrieved February 10, 2010).

[5] The sequence of the human DNA is stored in databases available to anyone on the Internet. The U.S. National Center for Biotechnology Information (and sister organizations in Europe and Japan) house the gene sequence in a database known as GenBank (Benson et al., 2007), along with sequences of known and hypothetical genes and proteins. Other organizations present additional data and annotation and tools for visualizing and searching it.

**6–Public II**

The first shift occurred in 1987, when a majority of scientists started to perceive the HPG as beneficial. Paul Berg described the change to the new regime as follows: "What is different, however, is how biologists view the project […]. There has been an enormous change in thinking about the project. […] [Earlier] we could hardly get to the science because of the ominous views people had about the project. I think now everyone agrees this is a worthwhile project, and we can get on to talking about how one might go about it in the most cost-effective and scientifically effective way." (Paul Berg, cited in Roberts, 1987b/237). Between 1987 and 1992, several relevant steps that brought genome sequencing forward, occurred. In 1989, PCR/STS was developed as a way to bring together different mapping techniques that had seemed incompatible, in order to facilitate cooperation among labs. It made traditional physical mapping obsolete (Roberts 1989b/245; Olson et al., 1989; Jordan and Lynch, 1998). In 1990, three groups developed capillary electrophoresis, and in the same year, Lipman and colleagues (NCBI) published the algorithm BLAST for aligning sequences (Roberts et al., 2001).

After 1992, other agencies outside the U.S. took on prominent roles, foremost Britain's Wellcome Trust (Sanger Center (UK), opening in 1993), and in October and December 1992, U.S. and French teams completed the first physical maps of chromosomes and the genetic maps of the mouse respectively. The year of 1993 can be considered as the "tipping point," defined as when the levels of development and commitment from various parties at which the momentum for the HGP became unstoppable (Gladwell, 2002). In January 1993, Walter Gilbert remarked that "today, there are ten-fold more [markers], and the role of genetic information is ten-fold more obvious to everybody." (Walter Gilbert, cited in Anderson, 1993/259: 300). Consequently, in October 1993, Francis Collins, head of NCHGR, requested more money to pursue genome research, on the basis that the budget had not increased as fast as the project's creators recommended. The combined NIH and DOE budget remained at roughly $165 million on 1992, when it should have been $219 million based on the planning in late 1980's and adjusted for inflation. Should they not increase the budget, Collins argued this would imply delayed medical benefits as well as loss of U.S. biotechnology competitiveness (Roberts, 1993/262).

In 1995, NHGRI began to accelerate the effort, funding six pilot projects in high-volume sequencing. Another turning point came in 1998 when Robert Waterston at Washington University in St. Louis, funded by NHGRI, and his collaborator John Sulston of the Sanger Centre near Cambridge, U.K., funded by the Wellcome Trust, announced that they had deciphered the complete genome (97 million bases) of the nematode, *Caenorhabditis elegans*.

Meanwhile, at the Institute for Genomic Research, Venter was perfecting a faster 'whole-genome shotgun' approach. He wowed to produce the complete genome of the bacterium *Haemophilus influenzae* (1.8 million bases long) using this technique, at record speed. And in May 1998, he dropped the bomb: backed by PE Corporation of Norwalk, Connecticut, he announced that Celera Genomics would sequence the entire human genome by 2001, using this whole-genome shotgun method.

**7–Private III: Venture-ing**

Venter's Institute for Genomic Research (TIGR) was soon joined by other biotechnology companies that competed directly with the publicly funded Human Genome Project. From 1992 onwards, genome scientists in universities found venture capitalists hammering on their doors (Cook-Deegan, 2000; 2004; Sulston and Ferry, 2002). Prospects for attracting private capital had changed dramatically in the five years since the Human Genome project was first outlined and the two years since its official start. In 1990 already, a symposium devoted to solicit interest among pharmaceutical firms, organized by Craig Venter and Walter Gilbert, drew a respectable audience (Cook-Deegan, 1994). Two years later, new created firms, joint ventures and other private companies became more and more attracted by the new potentials, comforted by the feeling that the business of sequencing was on a good track. Several small biotechnology firms redirected their efforts towards mapping and sequencing DNA, several new firms were founded (including three of the "big four", Human Genome Sciences, Incyte, and Millennium) (Cook-Deegan et al., 2000). As a result, the venture capital community was "getting very excited, all the pieces are coming together", as a venture capitalist has put it (Mark Levin, cited in Anderson, 1993/259: 301). The Human Genome Project at that time was perceived to be moving more quickly than anyone expected initially and was blessed by the characteristic "bubble mood" (Gisler and Sornette, 2009).[6] People were extrapolating by anticipating the development of therapeutics in short order, expectations that proved to be utterly exaggerated and removed from reality, as reviewed recently by Helen Pearson in Nature (2009/460). The HGP was now evolving from a public to a joint private/public effort, exemplifying the role that small entrepreneurial firms supported by venture capital play in the innovation process (Lamoreaux and Sokoloff, 2007). In fact, private funding reached rough parity with government and nonprofit funding in 1993 in the United States. Ever since, private genomics research funding has risen even faster (Cook-Deegan et al., 2000). One reason for the newly observed openness of the scientists involved in 1992/93 towards private funding might be the fact that the NIH and DOE budget remained relatively low (lower than announced initially, see above), and thus provoked delays in the effective support of ongoing research. Researchers, as a consequence, were more open to private investigators. The years 1992/3 was furthermore characterized by a shift from mapping to sequencing. Because industry is best at that kind of factory-like production, academic-industrial partnerships thus made sense. It was in any case a sign that the genome project was indeed succeeding.

Yet, during all this turmoil concerning the public-private rivalry, it seems to have been forgotten that the public HGP also contracted with private firms in order to get better, i.e. faster, machines to carry out the sequencing. It was Watson himself who, in 1992, argued that the benefits of industrial participation far outweighed the potential drawbacks. Academics had launched the project and were well on their way to finishing genetic maps. However, Watson saw the time had come to move to large scale sequencing, and industry was best at that kind of factory-like production. Academic-industrial partnership thus made sense, since the technology was already being developed in university labs. Watson himself helped establish a company, from a collaboration between researchers at Cold Spring Harbor Laboratory, which Watson directed, and Brookhaven National Laboratory, with the goal of developing high-speed sequencing technology (Anderson, 1993/259).

As another example, in March 1992, Walter Gilbert joined University of Utah geneticist Mark Skolnick in a company called Myriad Genetics Inc., where he still serves as vice chairman of the board. Myriad Genetics, founded in 1991, was devoted to developing cancer therapies by tracing genes turned up by the Human Genome Project. The company was funded by Eli Lilly and Co. and the investment banking firm Spencer Trask Inc. (Anderson, 1993/259). This case is characteristic of the sentiment of the time shared by venture capitalists on the attractiveness of investing into a formerly unknown "big" science project. Many among the venture capitalists were interested in financing projects connected to genome research, namely developing therapeutics (e.g. Mercator Genetics Inc.), while others were interested in focusing on the sequencing process itself.

The attitude toward the HGP had not only changed on Wall Street, but also among scientists themselves. While an upheaval occurred among them in 1987 when Gilbert announced his private going, a few years later, quite a few scientists were ready to join newly established joint ventures. Eric Lander, who in 1992 directed the single largest genome grant (a $24 million over five years) to map the entire human genome, still contended that genomic maps were basic infrastructure, and must thus be universally and freely available. However, even he agreed upon the idea of a company using public available maps to study particular genetic models of diseases (Anderson, 1993/259; Sulston and Ferry, 2002).

---

[6] "It took 4 years to obtain the first billion [base pair mark ] and 4 months to get the second billion. […] The goal for completing the working draft has not changed since it was first announced: 90% coverage of the euchromatic [informative] portion of the human genome sequence." (Francis Collins, cited in Roberts, 2000/287: 2396).

In 1999, the official HGP started itself to buy sequencing machines from Applied Biosystems Inc., the company which supported Craig Venter's Celera and which in fact initially developed machines for Celera only. This caused quite a stir at Celera, leading ABI to promise Celera a priority treatment (Shreeve, 2004). These sequencing machines, after all, were the descendents of the first automated sequencing machines developed by Leroy Hood and colleagues at Caltech in 1986, a publicly funded endeavor (Lewin, 1986a/233).

At the organizational level, the actors of the public program wasted no time in increasing the pace and in reorienting their schedule in an attempt to win the race. Changes were indeed very much needed, as a report reviewing the development of the public HGP insisted (Koonin, 1998). Francis Collins thus announced new goals for the public project in September 1998, six months after Venter's surprise announcement (Marshall, 1998/281). First, the consortium would complete the entire genome by 2003, two years ahead of schedule, but also two years behind the date announced by Venter. And, in a dramatic departure from previous philosophy, the project would produce a 'rough draft,' covering 90% of the genome by the spring of 2001. Scientists were clamoring for the data even in rough form, Collins said by way of explanation. Yet he also admitted that producing a rough draft and making it public was a strategic move to undercut any patent position Celera or other businesses might claim.

**8–The biotech financial bubble**

If we are correct that a bubble spirit was indeed developing in the social component of the HGP, there should be some observable signature of it in the financial markets. Indeed, new biotech companies dedicated to genomics, as well as established pharmaceutical firms positioned to exploit drug applications resulting from genomics, should have drawn high demand, as investors are often attracted by promises of great future incomes. High demand in turn pushes prices up. If the bubble spirit was active, the public-private race should have led to a kind of positive feedback, in which (i) the higher the belief in future gains, the higher the demand, (ii) the higher the demand, the higher the price, (iii) the higher the price, the higher the valuation of biotech companies, (iv) the higher the valuation of biotech companies, the more attractive and powerful they become, (v) the more attractive and powerful, the higher the demand … leading to an accelerating price spiral.

In order to test this hypothesis, we analyze the Amex Biotechnology Index (^BTK) and the Nasdaq Composite indices from Jan. 1997 to June 2002. The Amex Biotechnology Index is designed to measure the performance of a cross section of companies in the biotechnology industry that are primarily involved in the use of biological processes to develop products or provide services. The index is equal-dollar weighted, designed to ensure that each of its component securities is represented in approximate equal dollar value. Launched in 1971, the Nasdaq Composite Index is broad based and includes today over 3,000 securities, mainly in so-called new technology sectors, i.e, it includes the ICT (Internet-Communication-Technology) as well as the Biotech sectors. It is calculated under a market capitalization weighted methodology index and includes mainly U.S. firms listed on the Nasdaq Stock Market (with some exceptions).

Figure 1 shows the Biotech index over the time interval from Jan. 1997 to June 2002. Its inset shows the same data magnified from June 1998 to April 2000. One can observe an almost quadrupling of the index from 1998 to the peak occurring in early March 2000. Also notable is the fact that this quadrupling developed as an accelerated growth that can be termed "super-exponential", to stress the fact that the growth rate grew itself as the price increased. Recall that a constant growth rate qualifies just an exponential growth. Here, for the Biotech index, the growth is super-exponential, which means that investors expect for instance 10% return over the first 6 month period, then 20% over the next period, then 40%, then 80% … which is clearly unsustainable!

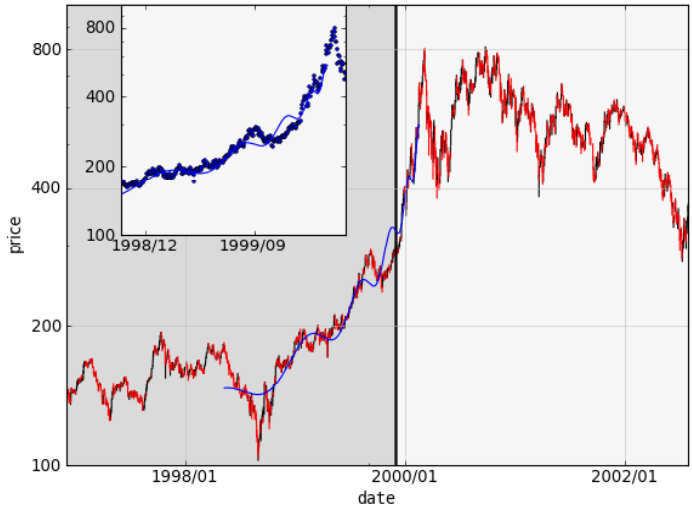

Fig. 1: Amex Biotechnology Index (in logarithmic scale) from Jan. 1997 to June 2002. The inset shows the same data magnified from June 1998 to April 2000. The vertical line indicates the time (30 Nov 1999) when the Biotech index disconnects and shoots up until the crash in early March 2000, leaving the Nasdaq index largely behind. The oscillating continuous lines in the main figure and inset correspond to the calibration of equation (1) to the Biotech index up to the peak. Note the upward curvature in this log(price) versus time, which qualifies a super-exponential accelerating price, qualifying a bubble.

Such super-exponential growth is our technical definition of a bubble, according to the methodology developed over the past 15 years in many papers and books by our group. We refer to the broad overviews (Johansen et al., 1999; Johansen and Sornette, 2006; Sornette, 2003; Sornette and Johansen, 2001; Sornette and Zhou, 2006; Jiang et al., 2010). In short, the methodology is based on the hypothesis that positive feedback on the growth rate of an asset's price by price, return and other financial and economic variables leads to faster-than-exponential (power law hyperbolic) price growth. The signature of positive feedbacks at work during a bubble is quantitatively identified in a time series by a faster-than-exponential power law component, and by the existence of increasing low frequency volatility, these two ingredients occurring either in isolation or simultaneously with varying relative amplitudes. A convenient mathematical representation has been found to be the existence of a power law growth decorated by oscillations in the logarithm of time. The simplest mathematical embodiment is obtained as the first order expansion of the log-periodic power law (LPPL) model:

(1)    $\ln P(t) = A + B |t - t_c|^\alpha + C |t - t_c|^\alpha \cos[\omega \ln |t - t_c| + \phi] + \varepsilon(t)$ ,

where $P(t)$ is the price of the asset, t is time and $\varepsilon(t)$ is a noise residual. There are seven parameters in this nonlinear equation, but two ($\alpha$ and $\omega$) stand out in their role for qualifying a bubble regime. Extensive tests have led to the hypothesis that the LPPL signals are excellent diagnostic tools of the existence of a bubble (see for instance, Sornette et al., 2009). The parameter $t_c$ represents the time at which the bubble ends, either in a crash or in a less-dramatic leveling off of the growth leading to a change of regime.

Figure 1 shows clear evidence of such (log-periodic) power law (super-exponential) growth of the Biotech index from Jan. 1998 to 2000, as exemplified by the overall upward curvature in this log(index) as a function of time. Figure 2 shows the Nasdaq index over the same period. The results are qualitatively and quantitatively similar. In particular, the super-exponential behavior is strikingly analogous to that observed for the Biotech index over the same period.

But was the biotech bubble illustrated in figure 1 really fueled by the HGP? Or could it be that the Biotech sector was in fact more driven by the ICT sector, the dot.com frenzy pushing with it all companies with a technological flavor to the sky? This second scenario is quite plausible since the ICT sector witnessed an extraordinary bubble from 1995 to 2000, which ended with a dramatic crash in April 2000 (Johansen and Sornette, 2000), as shown in figure 2. And the Biotech sector also crashed at the same time, as shown in figure 1.

To disentangle the HGP factor from the ICT factor with respect to their respective potential impact on the Biotech sector, we express the Biotech index "in currency units" of the Nasdaq index, i.e, we write

(2)     Biotech-in-Nasdaq(t) = Biotech-index(t) / Nasdaq(t) .

Taking the ratio of the Biotech and Nasdaq indices amounts to constructing a proxy for the HGP factor that we are trying to identify. This ratio has a clear economic meaning: it amounts to study the value of a portfolio that buys the Biotech sector and "shorts" (sells) the Nasdaq index. Equivalently, this Biotech-in-Nasdaq(t) ratio views the Nasdaq index as the currency used to purchase the Biotech index. In this way, we shortcut any influence of the U.S. dollar and directly extract the component of the bubble in the Biotech sector not present in the Nasdaq composite index. This approach is particularly well-suited to remove the influence of monetary policy as well as international influence on the value of the reference currency, which can play a big impact on the analysis of anomalous market regimes (Zhou and Sornette, 2005).

If both Biotech and Nasdaq indices move more or less in synchrony (in econometric jargon, this is often referred to as "co-integration" (Engle and Granger, 1987)), we should expect the Biotech-in-Nasdaq(t) to be more or less flat and noisy, which is the case before November 1999, as shown in figure 3. However, from the end of November 1999, indicated by the vertical line in figure 3, the Biotech-in-Nasdaq(t) shoots up until the peak on 7 March 2000 followed by the crash. Clearly, in the last three months of the unfolding of the two bubbles, the Biotech index took a life of its own, accelerating even faster than the Nasdaq index. The aftermath of the March 2000 crash is also strikingly different: the Nasdaq index does not recover over the period shown here until much after mid-2002, while the Biotech index strikingly recovers after the large crash and breaks its previous record in much less than a year, continuing its ascension till the end of 2001.

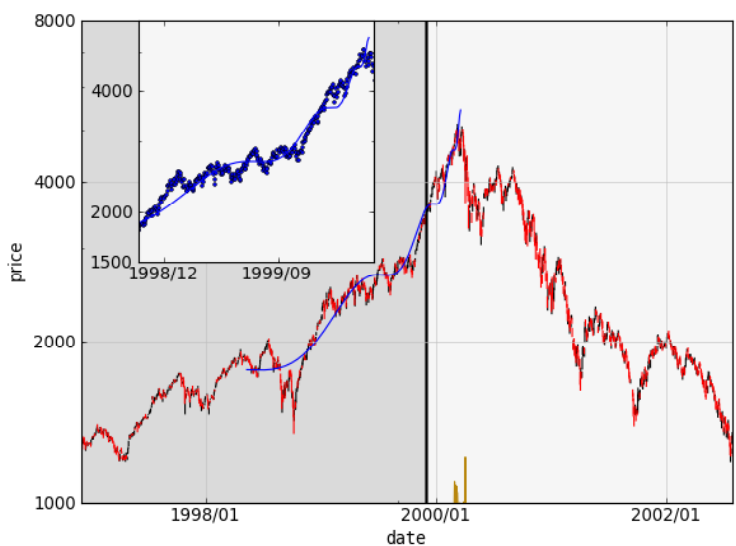

Fig.2: Nasdaq Composite indices (in logarithmic scale) from Jan. 1997 to June 2002. The inset shows the same data magnified from June 1998 to April 2000. The vertical line indicates the time 30 Nov 1999 (see Figs.1 and 3). The oscillating continuous lines in the main figure and inset correspond to the calibration of equation (1) to the Nasdaq index up the peak. Note the upward curvature in this log(price) versus time, which qualifies a super-exponential accelerating price, qualifying a bubble.

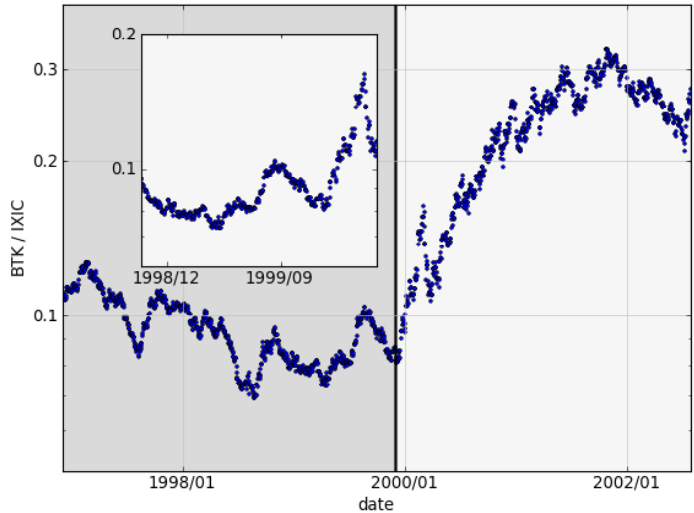

Fig.3: Biotech-in-Nasdaq(t) (= Biotech-index(t) / Nasdaq(t)) defined by equation (2) (in logarithmic scale) as a function of time from Jan. 1997 to June 2002. The inset shows the same data magnified from June 1998 to April 2000. The vertical line indicates the time 30 Nov 1999 (see Figs.1 and 2).

**9–Upshot**

In April 2000, the Subcommittee on Energy and Environment of the Committee on Science of the U.S. House of Representatives conducted hearings on the status and benefits of genome sequencing in the public and private sectors. Robert Waterston, director of the HGP sequencing center at Washington University, St. Louis, pointed to fruitful data sharing between the HGP and the private sector. Examples among others included collaborations led by the pharmaceutical company Merck to develop partial sequences identifying genes. These efforts showed that, despite the public-private race and the war rhetoric, sharing of data was finally perceived by all parties as a worthwhile endeavor in order to increase knowledge and ensure future discoveries.

Behind the scenes, Ari Patrinos of DOE played the mediator, and finally brokered a truce under which both groups would announce their drafts at the same time, thereby sharing the glory. Venter would still not deposit his data in GenBank, as the consortium wanted, but promised to publish his findings in accordance with the terms of the 1996 "Bermuda Statement," by releasing new data annually (in contrast, the public HGP released its new data daily). Unlike the publicly funded project, though, he would not permit free redistribution or commercial use of the data (Human Genome News, 1996/7/6). Eventually the HGP and Celera did manage to publish simultaneously their results, however in separate journals (Nature and Science respectively). And Venter finally conceded that the public data had been useful in his own work.

A 'rough draft' (not the full sequencing) of the genome was finished in 2000 (announced jointly by then U.S. president Bill Clinton and UK Prime Minister Tony Blair on June 26, 2000). Ongoing sequencing led to the announcement of the essentially complete genome in April 2003, two years earlier than initially planned. In May 2006, another milestone was passed on the way to full completion of the project, when the sequence of the last chromosome was announced. According to the definition employed by the International Human Genome Project, the genome has been completely sequenced by the end of 2003. However, there are still a number of regions of the human genome for which the project can be considered unfinished.

The fact that the project came to an end earlier than planned can thus be attributed mainly to the public/private competition, and not so much to the intrinsic increased speed of the underlying technology involved in sequencing, as has been argued on various occasions (e.g. The Department of Energy and the Human Genome Project Fact Sheet, from

[www.ornl.gov/sci/techresources/Human_Genome/project/whydoe.shtml](www.ornl.gov/sci/techresources/Human_Genome/project/whydoe.shtml); retrieved March 3, 2010). Rather, the competition increased efficiency, for the benefit of the project, by forcing the public effort to take more risks, leading to accelerated results that, in turn, helped the private initiative. Figure 4 provides a synoptic measure of the development of the HGP, by showing the number genomic patent applications per year after 1985. For all types of patents, the peak followed by a rather fast decay occurring in 2000 or 2001.

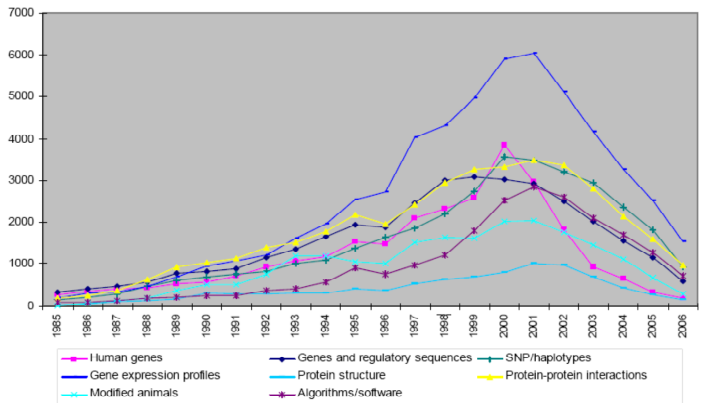

Figure 4: Overview of genomic patent applications after 1985. Source: Aurora Plomer and Peter Taylor, ESRC Complexity Se Nasdaq Composite indices minar Series, 26th November 2008.

**10–Discussion**

We have presented a detailed history of the development of the Human Genome Project from 1986 to 2003. This development is characterized by a formidable competition between the initially public project and several private initiatives, which became more and more prominent, so much as to force the former to adapt and change drastically its strategy. The explosion of interests and commitments from the private sector and from venture capitalists that continued till the completion of the project is the consequence of great expectations on commercial applications in drugs and medicine that could result from the sequencing and mapping of the entire human genome. The race and mutual interactions between the public and private HGP sustains the hypothesis that strong social interactions between enthusiastic supporters of the HGP weaved a network of reinforcing feedbacks that led to a widespread endorsement and extraordinary commitment by those involved in the project. This thus supports the social bubble hypothesis (Gisler and Sornette, 2009; Sornette, 2008).

But one could argue that the evidence presented here does not describe a bubble, but just the dynamics of a project based on rational expectations. Thus, it is worthwhile to briefly discuss whether or not the great anticipations on the commercial and medical applications of the HGP turned out to be fulfilled and on what time scales. As a matter of fact, now that the human genome has been sequenced almost completely, there is still little understanding of how genes actually work. Having the complete gene set on the table, the knowledge of the genetic map and sequence is now considered by experts to be only a starting point for future research in biology and medicine. It is now widely recognized that it will take decades to exploit the fruits of the HGP, via a slow and arduous process aiming at disentangling the extraordinary complexity of the problem (Pearson, 2009/460).[7] In this sense, the HGP illustrates vividly the "social bubble" hypothesis, according to which investors and actors develop extraordinary over-optimistic expectations of short-term applications during the development of a project, making them take risks that would not be justified by a standard cost-benefit analysis in the presence of huge uncertainties over long-time scales. It is the effect of social interactions and amplification that created the atmosphere in which the HGP bubble was catalyzed and could blossom (Sornette, 2008).

---

[7] See also e.g. Allan Bradley, Director of The Wellcome Trust Sanger Institute[9], Cambridge, UK, stating that "We shouldn't expect immediate major breakthroughs but there is no doubt we have embarked on one of the most exciting chapters of the book of life." (March 2004; [http://www.bbc.co.uk/dna/h2g2/A1091323#back9](http://www.bbc.co.uk/dna/h2g2/A1091323#back9); retrieved February 10, 2010).

Coming back to the issues raised in the introduction on the role of government as a public entrepreneur and venture capitalist for long-term very risky projects, we are led to conclude that the competition between the public and private sector actually played in favor of the former, since its financial burden as well as its horizon were significantly reduced (for a long time against its will) by the active role of the later. The fact that a social as well as financial bubble developed during the course of the HGP helped tremendously in this respect. This supports our hypothesis that social bubbles are essential carriers for pushing segments or even sometimes the whole of society to invest considerable efforts in very risky endeavors that brings enormous rewards only decades later, that is, after many capital investments have been lost on the short term. We go as far as suggesting that the government and public agencies were lucky in playing on the HGP bubble. This suggests that governments can take advantage of the social bubble mechanism to catalyze long-term investments by the private sector that would not otherwise be supported. Social bubbles thus provide a mechanism for aligning the apparently incompatible incentives of the private sector, that privileges (perceived) low-risk investments providing short-term returns, with the long-term social benefits of basic research for scientific and technical knowledge.

While there is little to show in terms of progress in medical diagnosis and treatment, in pharmaceutical development, in agriculture, and in other industrial sectors, the HGP catalyzed enormous technological progresses in DNA-based methods. As shown in figure 5, the cost of sequencing and mapping underwent an astonishing decrease. Actually, announced by Complete Genomics, a startup based in Mountain View, CA, a complete human-genome sequence (not a full genome sequencing!) can soon be ordered for $5,000, thanks to a new sequencing service. Such a stunning price drop may completely change the way human-genomics research can be carried out. A $5,000 genome would enable new studies to identify rare genetic variants linked to common diseases, and it could open up the sequencing market to diagnostic and pharmaceutical companies, making genome sequencing a routine part of clinical drug testing (see http://www.technologyreview.com/biomedicine/21466; retrieved March 3, 2010). Another illustration is the recent publication of a draft of the sequence of the giant panda genome with 2.25 gigabases, using so-called next-generation sequencing technology (Li et al., 2010). This work provides a foundation for comparative mammalian genetic research, and many usher novel applications. The fruits of the HGP are thus progressively coming, almost a decade after completion.

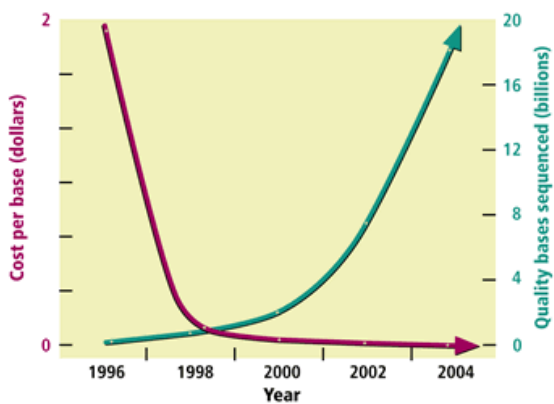

Figure 5: Illustration of the increasing efficiency measured by the fast decrease of the cost per base and the concomitant accelerating pace of sequencing; retrieved from www.ornl.gov/sci/techresources/Human_Genome/project/whydoe.shtml; June 1, 2009.

The HGP is initiating other initiatives, based on the recognition that there is much that genomics cannot do, and that "the future belongs to proteomics," according to Stanley Fields (researcher at Howard Hughes Medical Institute, and Adjunct Professor of Microbiology at the University of Washington School of Medicine, Seattle). Proteomics means the characterization of the entire array of proteins encoded by our genes. This is a huge task as different types of cells in the human body each have a different set of proteins, different protein structure and function can be modified in many ways, such as phosphorylation or glycosylation and, a single gene can encode for multiple proteins. All these possibilities result in a proteome that is an order of magnitude more complex than the genome, according to Fields as reported by Haroon Ashraf in The Lancet (2001/357: 531–2). Present efforts include searches for protein's involvement in diseases and its potential for a drug target and classifications of all the proteins and their [amino-acid] sequences. Will a new era emerge, that will promote a social proteomics bubble? The present work may help in understanding the necessary ingredients, the pros and cons, and the consequences.

**Acknowledgements**

We are grateful to Robert Cook-Deegan, Durham, NC, for providing us with the data collection on the money spent for biotechnology endeavors.

| U.S. Human Genome Project Funding | | | |
| --- | --- | --- | --- |
| ($Millions) | | | |
| FY | DOE | NIH* | U.S. Total |
| 1988 | 10.7 | 17.2 | 27.9 |
| 1989 | 18.5 | 28.2 | 46.7 |
| 1990 | 27.2 | 59.5 | 86.7 |
| 1991 | 47.4 | 87.4 | 134.8 |
| 1992 | 59.4 | 104.8 | 164.2 |
| 1993 | 63.0 | 106.1 | 169.1 |
| 1994 | 63.3 | 127.0 | 190.3 |
| 1995 | 68.7 | 153.8 | 222.5 |
| 1996 | 73.9 | 169.3 | 243.2 |
| 1997 | 77.9 | 188.9 | 266.8 |
| 1998 | 85.5 | 218.3 | 303.8 |
| 1999 | 89.9 | 225.7 | 315.6 |
| 2000 | 88.9 | 271.7 | 360.6 |
| 2001 | 86.4 | 308.4 | 394.8 |
| 2002 | 90.1 | 346.7 | 434.3 |
| 2003 | 64.2 | 372.8 | 437.0 |
| Total | 1015.0 | 2785.8 | 3798.3 |

Table 1: The DOE and NIH genome programs set aside 3% to 5% of their respective total annual budgets for the study of the project's ELSI issues (retrieved from www.ornl.gov/sci/techresources/Human_Genome/project/whydoe.shtml, June 1, 2009). Slightly different figures for the years 1988–1991 are given in Watson & Cook-Deegan, 1991.